\documentclass[conference]{IEEEtran}
\IEEEoverridecommandlockouts
\usepackage{cite}
\usepackage{amsmath,amssymb,amsfonts}
\usepackage{algorithmic}
\usepackage{graphicx}
\usepackage{textcomp}
\usepackage{xcolor}
\def\BibTeX{{\rm B\kern-.05em{\sc i\kern-.025em b}\kern-.08em
    T\kern-.1667em\lower.7ex\hbox{E}\kern-.125emX}}
\usepackage{tabularx,booktabs}
\newcolumntype{C}{>{\centering\arraybackslash}X} 
\usepackage{lipsum} 

\begin{document}

\title{Artificial Intelligence-based Eosinophil Counting in Gastrointestinal Biopsies}

\author{Harsh Shah, Thomas Jacob, Amruta Parulekar, Anjali Amarapurkar, Amit Sethi}

\maketitle

\begin{abstract}
Normally eosinophils are present in the gastrointestinal (GI) tract of healthy individuals. When the eosinophils increase beyond their usual amount in the GI tract, a patient gets varied symptoms. Clinicians find it difficult to diagnose this condition called eosinophilia. Early diagnosis can help in treating patients. Histopathology is the gold standard in the diagnosis for this condition. As this is an under-diagnosed condition, counting eosinophils in the GI tract biopsies is important. In this study, we trained and tested a deep neural network based on UNet to detect and count eosinophils in GI tract biopsies. We used connected component analysis to extract the eosinophils. We studied correlation of eosinophilic infiltration counted by AI with a manual count. GI tract biopsy slides were stained with H\&E stain. Slides were scanned using a camera attached to a microscope and five high-power field images were taken per slide. Pearson correlation coefficient was 85\% between the machine-detected and manual eosinophil counts on 300 held-out (test) images.    
\end{abstract}

\section{Introduction}

Eosinophils are the granulocytes that are formed inside the bone marrow. From bone marrow after maturation, they migrate to thymus, uterus, mammary glands, and the gastrointestinal (GI) tract. These are the normal residing sites of eosinophils in our body.~\cite{b1} They actively participate in protective mechanism against parasites and allergens by degranulating inflammatory mediators such as leukotrienes, vasoactive polypeptides and interleukins. An increase in their number is, therefore, indicative of an underlying condition, which can provide vital context for the management of diseases associated with these organs. For instance, eosinophils in GI tract are found in the lamina propria layer of mucosa.~\cite{b2} Eosinophils in GI biopsy normally shows orange coloured coarse granular cytoplasm and a bilobed nucleus. Normally eosinophils are present in the GI tract of healthy individuals.~\cite{b4} When the eosinophils increase than its usual amount in the GI tract, patient gets varied symptoms depending upon the site of involvement. Their presence in the GI tract varies in number based on various factors like geographical variation, age group, food allergies, type of food consumption by the populations, environmental hygiene and sanitation, any parasitic infestation and drug allergies. Eosinophils are increased secondary to diseases commonly like Celiac disease, Ulcerative colitis, Chron’s disease, Malignancies, Eosinophilic syndromes and drug induced eosinophilia.~\cite{b3}

Due to varied symptoms, the clinician finds difficulty in diagnosing eosinophilia. As there is a delay in the diagnosis, it can lead to progression of chronic diseases. Histopathology is helpful in diagnosing the eosinophilia, which is a benign inflammatory disorder. Timely diagnosis can benefit the patient and treatments with steroids, Leukotriene antagonists, Anti- IgE and Anti- IL-5 can be given. Histopathology is the gold standard in the diagnosis for this condition. Thus, there is a need of counting eosinophils in the GI tract biopsies as it is a common, often subclinical, underdiagnosed, and diagnostically challenging entity.~\cite{b5}    
            
Artificial intelligence (AI), especially in the form of deep convolutional neural networks, can be used for detecting objects in images. For detecting eosinophilia, we trained and tested deep neural networks on carefully annotated images of GI tract histopathology and tested the models on held-out images.~\cite{b6} We hope that when this technology is deployed in the clinics after further validation, it will take comparatively lesser time to carry out the function done manually, be more effective than relatively lesser experienced pathologists, and work without fatigue.

\subsection{Aims and Objectives}

We summarize the objectives of the study as follows:
 
\begin{enumerate}
\item To train and test AI models for recognizing and counting eosinophils in histological slides.
\item  To study correlation of eosinophilic infiltration counted by AI and manual method.

\end{enumerate}

Towards this end, we carefully annotated eosinophils in 400 cases of GI tract biopsies, separated the images into training and testing at the patient-level, trained models based on the UNet~\cite{b8} with data augmentation, and examined various metrics, including correlation between actual and predicted eosinophil counts for the test images.

\section{Materials and Methods}

\subsection{Study design}

The design of the study can be summarized as follows:

\begin{enumerate}
    \item \textbf{Study description:} Non-randomised prospective comparative study.
    \item \textbf{Place of study:} Department of Pathology, Lokmanya Tilak Municipal Medical College and Sion Hospital, Mumbai.
    \item \textbf{Duration of study:} 14 months (January 2021-February 2022).
    \item \textbf{Sample size:} 400.
    \item \textbf{Inclusion criteria:} (1) All gastro-intestinal tract (GIT) biopsy samples received in Department of Pathology, (2) all images of high-power fields ranging from 0 to 255-pixel intensities for eosinophil prediction by AI.
   \item \textbf{Exclusion Criteria:} Sections thicker than three micrometres, (2) sections smaller in size than required, (3) slides with uneven sections taken by the microtome, (4) over-staining or under-staining with haematoxylin and eosin (H\&E) stain, (5) slides with air bubbles trapped after mounting, (6) blurred images of high-power fields, and (7) eosinophilic or basophilic colour imbalance in the images.
\end{enumerate}

\begin{figure*}[t]
\centering
\includegraphics[width=\linewidth]{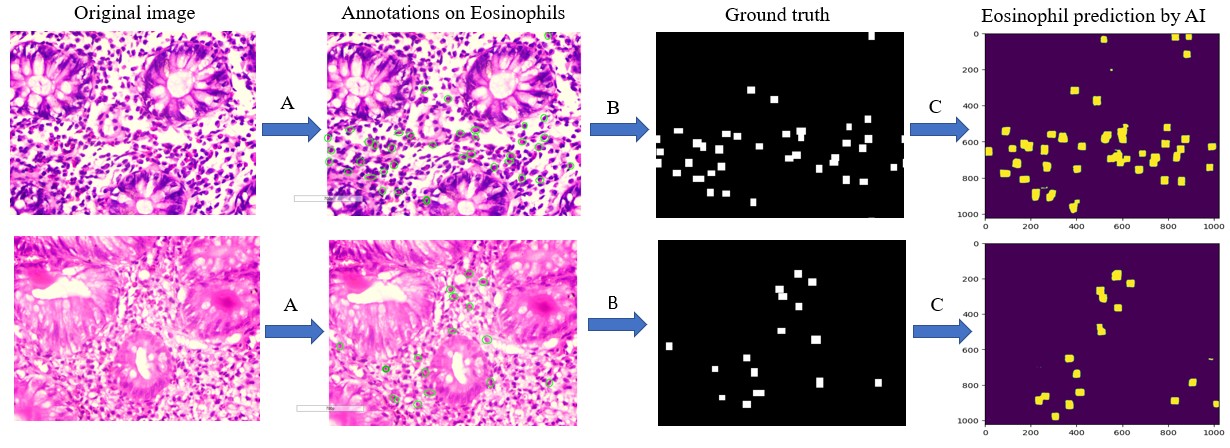}
\caption{Overall idea: Raw images of H\&E stained slides of GI tract biopsies, (taken using a camera attachment to a microscope), annotations of eosinophils (using Aperio ImageScope), ground truth map, and a sample prediction map produced by our method.}
\label{fig:overall}
\end{figure*}

\subsection{Study material}

All gastro-intestinal tract biopsy samples received in the Department of Pathology from January 2021 to February 2022 were included in this prospective study. The biopsy samples were taken for routine tissue processing. Paraffin blocks preparation and section cutting of three micrometres thickness was done using a microtome. Slides were stained with haematoxylin and eosin (H\&E). After diagnosis of the case, slides were taken for image scanning using a microscope with a camera port. For each slide, five such high-power fields were selected for taking images. 

\subsection{Annotation and software used}

For AI to recognize eosinophils, image annotation was done using Aperio ImageScope software. Images were fed to a convolutional neural network model for training and testing using python language and numpy, pytorch, and matplotlib libraries. This process is described in Figure~\ref{fig:overall}.

\subsection{Training details}

For training the AI model 100 cases were randomly selected and the remaining 300 cases were taken for testing and statistical analysis. The image data was standardized using color normalization, which helped in training the AI.~\cite{b7} Data was augmented using geometrical image transformations (flips and 90-degree rotations), and brightness augmentations. Annotations used to create mask images were converted from .xml files to binary mask using a python script. The samples for training were split for training (80\%) and validation (20\%). Model learns from the training data and validation shows evaluation of the model during it is training phase. All images are resized to 1024x1024 resolution for standardisation and divided by 255 to map them from 0 to 1 in the pixel intensity space and then fed to a UNet architecture.~\cite{b8} 

Loss function was performed with binary cross entropy and Dice losses combination to quantifies the difference between the prediction and ground truth.~\cite{b9} We used early stopping if there was no improvement in the validation Dice loss for 8 epochs. We used a learning rate scheduler to reduce the learning rate by a factor of 0.1 if there was no improvement in the validation Dice loss for 4 epochs. The initial learning rate was 0.0003 with Adam optimizer. We use the TensorBoard to monitor the loss. The output of the model is in the form of 1024x1024 single channel output. The output of the model is threshold at 0.8 or more than 0.8 to count it as an eosinophil. This cut-off was determined using validation. The function “connectedComponentsWithStats” from OpenCV was used to retrieve the count from the output. 

\subsection{Assessment of results}

AI-determined eosinophil counts were compared with the manual counts of the same cases which were already calculated by the co- investigator. For statistical analysis, we used Pearson correlation coefficient across cases.

\section{Results}

\subsection{Distribution of the cases}

A distribution of the cases is described below:

\begin{enumerate}
    \item Age distribution among the cases:
Among the 300 tested samples, majority of the samples were belonging to the age group of 23 to 32 years, that is 77 (25.6\%). 25 (8.33\%) samples were from patients of paediatric age group (<12 years). Median age was 34 years.

\item	Sex distribution among the cases:
   Among the patients, 59\% were male and 41\% were females. Thus, the Male: Female ratio is 1.4:1.
 
\item	Distribution according to Clinical symptoms:
   Abdominal pain was the most common (35\%) clinical presentation followed by chronic diarrhoea in 32\% cases. Only 1 (0.33\%) patient presented with constipation.

\item	Distribution according to Endoscopic findings:
   Endoscopic findings noted in 165 (55\%) patients were normal. Among the abnormal findings, mucosal ulceration was the common finding 79 (26.33\%) and only 2 (0.6\%) cases showed polyps.  

\item	Distribution according to biopsy site:
    Most number of biopsies were taken from Duodenum- 149 (49.66\%), followed by Ileum-35 (11.66\%).  
\end{enumerate}

\subsection{Eosinophil prediction}

\begin{figure}[htbp]
\centering
\includegraphics[width=\linewidth]{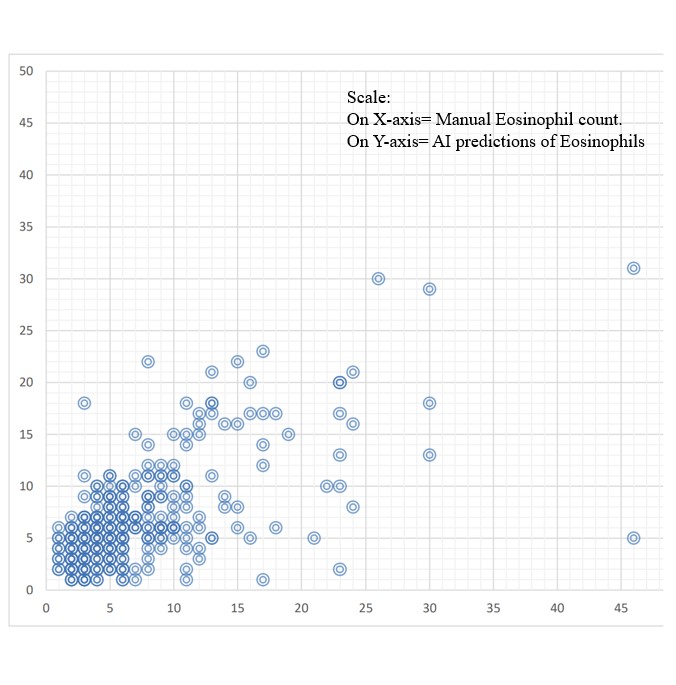}
\caption{Scatter plot to study the correlation between AI prediction of eosinophils with manual eosinophil count}
\label{fig:scatter}
\end{figure}

A qualitative sample of the results is shown in Figure~\ref{fig:overall}. A scatter plot to study the correlation between AI prediction of eosinophils with manual eosinophil annotation can be seen in Figure~\ref{fig:scatter}. An analysis of correlation between Eosinophils count by AI and manual method can be seen in Table~\ref{tab:correlation}.

\begin{table}[htbp]
\begin{center}
\caption{Correlation between eosinophils count by AI and manual method}
\label{tab:correlation}
\begin{tabular}{|c|c|c|}
\hline
Biopsy site  
& Correlation between & Interpretation \\ & eosinophil count by AI & (Correlation type) \\&  and manual method &  \\ & (Pearson correlation coeff.) & \\ 
\hline
Esophagus  & 0.41     & Medium strong positive     \\ 
Duodenum   & 0.55        &  Medium strong positive  \\ 
Ileum       & 0.84       &  Very strong positive     \\ 
Caecum      & 0.83       & Very strong positive    \\ 
Ascending colon       & 0.81      & Very strong positive      \\
Transverse colon  & 0.94     &  Very strong positive     \\ 
Descending colon   & 0.86       &   Very strong positive  \\ 
Sigmoid colon      & 0.98      & Very strong positive    \\ 
Rectum       & 0.9      & Very strong positive      \\ 

\hline
\end{tabular}
\end{center}
\end{table}

\begin{table*}
\caption{Comparison of prevalence of duodenal eosinophilia amongst our study and three other studies}
\label{tab:comparison}
\begin{tabularx}{\textwidth}{@{} l *{5}{C} c @{}}
\toprule
Study 
& Our study
 & Czyzewski T. et al.~\cite{b9} & Archila L. R. et al.~\cite{b10}
 & Catalano A. et al.~\cite{b11}
 \\ 
\midrule
Year of study   & 2022      & 2021         & 2021   & 2022 \\ 
Place & Mumbai, India  & Cincinnati, USA
        & Minnesota, USA
    & Columbia, USA
  \\ 
Number of cases      & 400       & 420        & 40  & 36  \\ 
Inclusion criteria      & All gastrointestinal tract biopsies received. 
       & Esophageal biopsies
         & Esophageal biopsies
  & Esophageal biopsies
 \\ 
AI model architecture      & U-Net architecture modification.
       & ResNet50 Deep convolutional neural network.
         & Nested individual Convoluted neural networks
   & U-Net architecture modification.  \\ 
Statistical analysis      & Pearson correlation coefficient
      & Sensitivity, Accuracy and positive predictive prevalence
         & Sensitivity and positive predictive prevalence
   & Accuracy with standard error
 \\ 
Final analysis      & Strong positive Eosinophil count correlation of
85\% (+/- 5\%) accuracy
       & Approximately 85\% accuracy
         & Very good accuracy-80-95\%

   & 91\% accuracy with difference of 8 eosinophils 
(+/- 1.5 SD)
 \\

\bottomrule
\end{tabularx}
\end{table*}

\section{Discussion and Prior Work}
      
Our study was performed for all gastrointestinal tract biopsies from the esophagus to the rectum. For image analysis, images of five non-overlapping high-power fields were taken for 400 cases. Annotations were performed on eosinophils (single layer annotations). For AI model, a modified U-Net architecture was used. Since the cases of esophagus and stomach were few, the correlation was medium strong positive. In cases of biopsies from other sites, the correlation was strong positive as the samples were more, hence the correlation improved. Overall accuracy was found to be 85\% (+/-5\%).

Previous studies on eosinophil counting were done only on esophageal biopsies. As AI is sensitive to the context, we believe that expanding that context to the entire GI tract was important. We summarize the previous studies below.

Czyewski T. et al performed their studies on esophageal biopsies. They used whole slide images (WSI) for eosinophil quantification. Total 420 whole slide images were used in the study. They used ResNet50 DCNN for Image analysis and quantification. In this model, eosinophilic infiltration in the form of patches was analysed, thus it becomes easy to quantify eosinophils in patches. So, the cropping of WSI into patches and downscaling these patches into inputs into 224x224 for AI training was done. They used sensitivity, accuracy and positive predictive prevalence for statistical analysis which showed 85\% accuracy. Accuracy turned out to be lesser as the images were downsized after cropping them, on downscaling, many eosinophils would have lost their differentiation as the quality of the image was compromised.~\cite{b10}
    
Archila L. R. et al performed their study on esophageal biopsies. Their sample size was 40 whole slide images. They used Nested individual CNN for eosinophil quantification. This model apart from eosinophil differentiation also had differentiation for collagen, intra-cellular oedema, lymphocytes, squamous nuclei. Thus, the layer wise (superficial squamous, basal layer and lamina propria) and cellular (eosinophils, lymphocytes and squamous cells) differentiation can give the output in the form of eosinophilic esophagitis (EoE) or non-eosinophilic esophagitis (Non-EoE). They used statistical tools such as Sensitivity and positive predictive prevalence. The model showed very good accuracy of 80-95\% with performance score-2.~\cite{b11}

Catalano A. et al studies on esophageal biopsies. Thirty six esophageal biopsies were used for AI training and testing. Like our study, they created the AI model by using a modified U-Net architecture. The output was evaluated using accuracy and standard error of the predictions. The accuracy was 91\% with difference of 8 eosinophils (+/-1.5 S. D.) The difference in AI predictions for eosinophils as compared to manual counts can be reduced by increasing the sample size with this AI model.~\cite{b12}

Table~\ref{tab:comparison} illustrates the comparison of prevalence of duodenal eosinophilia amongst our study and the other three studies.

\section{Conclusions}

Overall Pearson correlation of eosinophil count of the AI model in our study was found to be 85\% (+/-5\%) for the entire GI tract. The accuracy of this model can be improved by increasing the sample size of the biopsies which shows medium positive correlation to make it a strong positive correlation. Correlation of AI prediction for eosinophils with manual eosinophil count showed moderate to strong positive correlation. There is a need of counting eosinophils in the GI tract biopsies as it is a common, often subclinical, underdiagnosed and diagnostically challenging entity. Medications like steroids, leukotriene antagonists, Anti- IgE and Anti- Interleukin-5 have shown significant improvement in patients with eosinophilia.

\end{document}